# Chalcogenide perovskite BaZrS$_3$ thin-film electronic and optoelectronic devices by low temperature processing

Zhonghai Yu[a,1], Xiucheng Wei[b,1], Yixiong Zheng[c,1], Haolei Hui[b], Mengying Bian[b], Samyak Dhole[c], Jung-Hun Seo[c,*], Yi-Yang Sun[d], Quanxi Jia[c], Shengbai Zhang[e], Sen Yang[a,*], and Hao Zeng[b,*]

[a]MOE Key Laboratory for Nonequilibrium Synthesis and Modulation of Condensed Matter, Xi'an Jiaotong University, Xi'an 710049, China

[b]Department of Physics, University at Buffalo, the State University of New York, Buffalo, NY 14260, USA

[c]Department of Materials Design and Innovation, University at Buffalo, the State University of New York, Buffalo, NY 14260, USA

[d]State Key Laboratory of High Performance Ceramics and Superfine Microstructure, Shanghai Institute of Ceramics, Chinese Academy of Sciences, Shanghai 201899, China

[e]Department of Physics, Applied Physics & Astronomy, Rensselaer Polytechnic Institute, Troy, NY 12180, USA

[1] These authors contributed equally to this work.

* corresponding authors: junghuns@buffalo.edu; yangsen@mail.xjtu.edu.cn; haozeng@buffalo.edu

**Abstract:**

Owing to its superior visible light absorption and high chemical stability, chalcogenide perovskite barium zirconium sulfide ($BaZrS_3$) has attracted significant attention in the past few years as a potential alternative to hybrid halide perovskites for optoelectronics. However, the high processing temperatures of $BaZrS_3$ thin films at above 1000 °C severely limits their potential for device applications. Herein, we report the synthesis of $BaZrS_3$ thin films at temperatures as low as 500 °C, by changing the chemical reaction pathway. The single phase $BaZrS_3$ thin film was confirmed by X-ray diffraction and Raman spectroscopies. Atomic force microscopy and scanning electron microscopy show that crystalline size and surface roughness were consistently reduced with decreasing annealing temperature. The lower temperatures further eliminate sulfur vacancies and carbon contaminations associated with high temperature processing. The ability to synthesize chalcogenide perovskite thin films at lower temperatures removes a major hurdle for their device fabrication. The photodetectors demonstrate fast response and an on/off ratio of 80. The fabricated field effect transistors show an ambipolar behavior with electron and hole mobilities of 16.8 $cm^2$/Vs and 2.6 $cm^2$/Vs, respectively.

**Introduction**

Organic-inorganic hybrid halide perovskites have taken the center stage in photovoltaics research over the past decade [1-4], since Kojima *et al.* first introduced their application as a visible-light sensitizer in 2009 [5]. Many studies have further established their outstanding performance for optoelectronic applications, such as photodetectors [6, 7], light emitting diodes (LEDs) [8, 9], optical sensors [10, 11], and especially solar cells [12, 13]. However, the best performing halide perovskites for solar cells often contain the toxic element Pb [5, 14]. Furthermore, hybrid perovskites show inherent instability against moisture, heat, light illumination, and electric field. These disadvantages have severely impeded their commercial applications. A lot of effort has been devoted to enhancing the stability of organic perovskites, such as A-site cation engineering (*e.g.* replacing $MA^+$ by $FA^+$ and $C_S^+$) [15-17] and dimension reduction (*e.g.* using the two-dimensional Ruddlensden-Popper phase) [18-20].

To address the toxicity and stability limitations of hybrid perovskites, inorganic chalcogenide perovskites have been proposed as an alternative family of perovskite semiconductors for optoelectronics. In 2015, Sun *et al.* theoretically screened 18 $ABX_3$ chalcogenide perovskites for photovoltaics, with A as Ca/Sr/Ba, B as Ti/Zr/Hf, and X as S/Se [21]. Several $ABX_3$ chalcogenide perovskites were identified to have suitable band gaps and absorption properties for photovoltaics. As a prototypical chalcogenide perovskite, $BaZrS_3$ possesses a direct band gap of 1.8 eV, and strong near edge absorption [22-26]. Furthermore, $BaZrS_3$ has been shown to possess a stable perovskite structure against high pressure, moisture, and heat [23, 27]. However, little is known about its carrier transport properties as the majority of publications are on bulk powder or single crystals. We first reported the synthesis of $BaZrS_3$ thin films by high temperature (~1000 °C) sulfurization of $BaZrO_3$ films deposited via pulsed laser deposition (PLD) [28]. These films show *n*-type conductivity due to sulfur vacancies, with a carrier mobility of around 10 $cm^2/Vs$. Comparotto *et al.* fabricated $BaZrS_3$ thin films by co-sputtering of BaS and Zr targets, followed by a thermal treatment at temperatures varying from 650 to 1000 °C. It was reported that 900 °C was required for good crystallization. However, no transport properties were reported [29]. Gupta *et al.* synthesized $BaZrS_3$ thin films by sulfurization of $BaZrO_3$ films made by chemical solution deposition at 1050 °C [30]. Such high processing temperatures are incompatible with device fabrication, and could also lead to unintended high concentration of electron doping due to sulfur vacancy defects. Thus, lowering the processing temperature of chalcogenide perovskite thin films is a crucial step for establishing this family of materials as a viable competitor to hybrid organic perovskites for practical applications.

In this paper we report the synthesis of $BaZrS_3$ thin films at temperatures as low as 500 °C, which is comparable to the processing temperature of copper indium gallium selenide (CIGS) solar cells [31, 32]. This is achieved by changing the chemical reaction pathway from sulfurization of oxide perovskites to crystallization of pulsed laser deposited amorphous $BaZrS_x$ films. The films processed at low temperatures show stoichiometric single-phase orthorhombic perovskite structure with smooth surfaces. Elimination of excess sulfur vacancies leads to reduced carrier concentration to $\sim 4\times 10^{10}$ $cm^{-3}$. As a demonstration for the potential use of the BaZrS$_3$ thin films towards various electronic and optoelectronic applications, BaZrS$_3$ thin-film photodetectors and field effect transistors (FET) were fabricated. The photodetector measurements show photocurrent close to 2 orders of magnitude greater than dark current for the films annealed at 650 °C. The FETs with ionic liquid (IL) gating show ambipolar behavior, with an on/off ratio of $6.5\times10^3$ at gate voltage of 1 V and carrier mobilities estimated to be 16.8 cm$^2$/Vs and 2.6 cm$^2$/Vs for electrons and holes, respectively. Our study paves the way for the integration of chalcogenide perovskite thin films in optoelectronic devices.

**Experimental**

**Preparation of the BaZrS$_3$ PLD target**: BaZrS$_3$ powder was synthesized from BaZrO$_3$ powder following published procedure [24]. The pre-synthesized BaZrS$_3$ powder was cold-pressed into a pellet with a 25 T hydraulic press using a circular mold of 20 mm in diameter. Before pressing, a small amount of polyvinyl alcohol (PVA) was added to enhance the adhesion and stability of the pellet. The BaZrS$_3$ pellet was sintered at 1050 °C in Ar atmosphere for 2 hours, and used as target for PLD.

**Synthesis of BaZrS$_3$ thin films**: A 1 cm×1 cm (0001) sapphire wafer was loaded into the ultrahigh vacuum (UHV) deposition chamber of the PLD system equipped with a KrF pulsed excimer laser ($\lambda$=248 nm), at a base pressure of $1\times10^{-8}$ torr. The substrate holder was kept at temperatures of 500-700 °C during the deposition, with a laser repetition rate of 5 Hz and laser fluence of 5 J/cm$^2$. During the deposition, the substrate and target were both rotated at 30 rpm to ensure homogeneity of the films. The as-grown amorphous BaZrS$_x$ thin films were cooled down to room temperature in the UHV chamber and then transferred to an MTI two-zone tube furnace in a 2-inch diameter quartz tube for annealing under CS$_2$ carried by Ar gas flow at 12 standard cubic centimeters. The heating profile and a sketch of the setup are shown in the supporting information (SI) Fig S1. The as-deposited thin films were placed at the center of zone 2 downstream, with temperatures ranging from 500 to 900 °C and corresponding total annealing time ranging from 2 to 8 hours. The temperature of zone 1 was kept at 900 °C for complete decomposition of CS$_2$, as the decomposition rate of CS$_2$ is highly temperature dependent [33].

**Thin film characterizations:** The X-ray Diffraction (XRD) θ-2θ scans were performed using an X'pert Pro X-ray diffractometer operating at 1.6 kW (Cu Kα radiation). Raman spectra were obtained from a HORIBA Raman spectrometer working under 532 nm laser excitation. The scanning electron microscopy (SEM) images and energy dispersive X-ray spectroscopy (EDX) analysis were acquired from a Focused Ion Beam-Scanning Electron Microscope (FIB-SEM) – Carl Zeiss AURIGA CrossBeam with an Oxford EDS system. The Bruker AXS-Innova atomic force microscope (AFM) was used to obtain the surface roughness and thickness profiles. The absorption spectra were collected from a Cary series UV-Vis-NIR spectrophotometer measured from 400 nm to 800 nm.

**Device fabrication and Characterization:** The photodetector devices were fabricated by depositing the source and drain electrodes by electron beam evaporation of a 30 nm thick Au film through a shadow mask, with a deposition rate of 0.4 Å/s, using an AJA Dual ATC-Orion

UHV system with a base pressure of $1\times10^{-8}$ torr. The photodetector has a channel width of 1000 µm and a gap of 50 µm. The I-V curves were measured by a Keithley 4200-SCS Semiconductor Parameter Analyzer, in the dark and under illumination with an excitation wavelength of 532 nm and light intensity of 0.14 W/cm$^2$. The FET devices were fabricated by the same technique through a T-shaped mask. The FET has a channel length of 50 µm and a width of 1000 µm. The side gate is separated from the source and drain electrodes by a 150 µm gap. A drop of the IL Diethylmethyl(2-methoxyethyl)ammoniumbis(trifluoromethylsulfonyl)imide (DEME-TFSI), about 20 µL, was dispensed to cover the active area of the BaZrS$_3$ FET. The device was baked at 120 °C for 12 hours in a N$_2$ glove box to remove any moisture in the IL. The drain-source current $I_{DS}$ as a function of drain-source voltage $V_{DS}$ under different gate voltages $V_G$ and drain-gate capacitance $C_{DG}$ vs drain-gate voltage $V_G$ were measured by a Keithley 4200-SCS Semiconductor Parameter Analyzer.

**Results and discussions**

Conventional semiconductors and oxide/halide perovskites represent two ends of the spectrum in terms of the covalency-iconicity duality. Chalcogenide perovskites clearly belong to the ionic side of the spectrum, in stark contrast to conventional four-fold coordinated semiconductors. As a crude quantification of the effective ionicity/covalency, one may calculate the Pauling electronegativity ($\chi$) difference per anion [34]. For BaZrS$_3$, $\Delta\chi = \frac{[3\chi(S) - \chi(Ba) - \chi(Zr)]}{3} = 1.84$. Using the same methodology, $\Delta\chi$ is found to be 1.43 for AlN and 0.37 for GaAs. To put it into perspective, BaZrS$_3$ has a moderate band gap of 1.8 eV, close to that of GaAs; but its effective ionicity is even higher than that of wide gap AlN with a band gap of 6 eV. Possessing a moderate band gap with an exceptionally large ionicity puts BaZrS$_3$ and other chalcogenide perovskites in a unique category of defect tolerant semiconductors that may offer new opportunities for electronic and optoelectronic applications.

The high processing temperature of BaZrS$_3$ thin films in earlier studies was mainly attributed to the chemical reaction pathways with high activation barriers. As both group IIA and group IVB elements in perovskites have low electronegativity, their oxides are very stable and difficult to be converted into chalcogenides. In the approach of co-sputtering of BaS/Zr [29], the high processing temperature above 900 °C may also be related to the difficulty of the chemical reaction between BaS and Zr, as Zr is a refractory metal with a high melting point of 1855 °C. The residual oxides during the handling may further exacerbate the formation of BaZrS$_3$ using sulfur. Therefore, to reduce the processing temperature, a change of the reaction pathway is necessary. PLD has been used to deposit a wide range of compounds including oxides, nitrides and carbides [35-37]. PLD is also known to be able to keep the stoichiometry of the target material under optimal conditions [38]. Postulating that by PLD deposition of a stoichiometric BaZrS$_3$ compound target under optimal conditions, BaZrS$_3$ perovskite thin films can be obtained under milder conditions.

We found that the stoichiometry of the Ba:Zr cations of the as-deposited films is dependent on laser fluence of the PLD system. This effect can be understood as primarily due to the large difference in the cohesive energy of Ba vs. Zr (1.9 eV/atom vs. 6.25 eV/atom). Therefore, low laser fluence is expected to result in Ba-rich films, while very high laser fluence may result in Ba deficiency due to reverse sputtering. Under a narrow window of fluence of close to 5 J/cm$^2$, a 1:1 ratio of Ba:Zr was obtained, as shown in Fig S2(a), However, it was found that the as-deposited films were amorphous, even with an in-situ substrate temperature as high as 700 °C (Fig S3). It was further determined that these films are highly sulfur deficient, with a composition of BaZrS$_{2.4}$, as shown by the EDX spectroscopy results (Fig S2(b)), which perhaps contributes to their poor crystallinity.

To obtain stoichiometric, single phase BaZrS$_3$ in the orthorhombic perovskite structure, thermal annealing of the as-deposited amorphous BaZrS$_{2.4}$ films in CS$_2$ at various temperatures ranging from 500 °C to 900 °C were performed in a two-zone tube furnace. All of the BaZrS$_3$ films after annealing appeared dark red in color (see optical images in Fig 1(b)), except the one annealed at 500 °C, indicating strong visible light absorption. The 500 °C sample does show lighter color, suggesting incomplete crystallization. The sulfur concentration as measured by EDX is annealing temperature dependent, as seen in Fig S4. While as-deposited BaZrS$_{2.4}$ films are sulfur deficient, the composition of the films progressively approaches perfect stoichiometry of Ba:Zr:S=1:1:3 with decreasing annealing temperature. To investigate the annealing temperature dependent structural evolution of the BaZrS$_3$ films, XRD θ-2θ scans were performed for films annealed at 900 °C, 850 °C, 650 °C, 550 °C, and 500 °C, respectively, as shown in Fig 1(a). The peaks labeled with Miller indices indicate standard peak positions of BaZrS$_3$ with a Pnma structure, retrieved from the PDF card JCPDS 00-015-0327. It is found that all discernible peaks for the samples annealed from 500 to 900 °C match well with the reference. No additional peaks were observed, suggesting that the obtained films were of a single phase. Some diffraction peaks with low intensities were not clearly observed from films annealed at 550 °C and 500 °C, as a result of low crystallinity and small grain sizes at such low temperatures. As can be seen from Fig 1(b), the full width at half maximum (FWHM) of the (121) diffraction peak decreases from 0.62° to 0.45° with increasing annealing temperatures from 550 °C to 900 °C. The crystallite sizes as calculated from Scherrer's formula using the FWHM of the (121) diffraction peak decreases monotonically with decreasing annealing temperature. The lateral grain sizes, as imaged by AFM and SEM (shown in Fig 2(a)-(h) and Fig S5) also decreases monotonically from ~110 nm at 900 °C to 35 nm at 550 °C. It is interesting to note that the lateral grain sizes are several times larger than the vertical grain sizes. Understanding the detailed crystallization mechanisms in the future will help to obtain films with enhanced grain size and crystallinity.

The surface roughness of a semiconductor thin films plays an important role in determining the interface quality, carrier transport in junctions, and thus device performance [39, 40]. It is expected that a lower annealing temperature with smaller crystallite size should also possess lower surface roughness. The AFM top-view images in Fig 2(a)-2(d) and 3D topography images displayed in Fig 2(e)-2(h) show clearly that the crystallite size and surface roughness of the BaZrS$_3$ thin films decreases consistently with decreasing annealing temperature from 900 °C to 550 °C, respectively. The height profile measured along a scan length of 5 μm for different annealing temperatures is shown in Fig 2(i). The root mean square (RMS) of the surface roughness R$_q$ decreases by 20 folds from 10.5 nm at annealing temperature of 900 °C to 0.59 nm at annealing temperature of 650 °C, as seen in Fig 2(j). This trend can be explained by the consequence of a decrease in grain size with decreasing annealing temperature.

The room temperature Raman spectra of BaZrS$_3$ films annealed at 900 °C, 850 °C, 650 °C, and 550 °C are shown in Fig 3(a). Several peaks can be identified in the range of 50 cm$^{-1}$ to 500 cm$^{-1}$ and six of them are assigned to be $B_{1g}^1$, $A_g^4$, $B_{2g}^6$, $B_{1g}^4$, $B_{1g}^5$ and $B_{3g}^5$ vibrational modes calculated theoretically for BaZrS$_3$ with the Pnma structure [27]. The Raman spectra measured at low temperature (100K) are provided in Supporting Information Fig. S6. In addition, we measured the Raman spectra at higher wavenumbers between 1000 cm$^{-1}$ and 2000 cm$^{-1}$ for all the films to investigate the possibility of carbon contamination resulting from CS$_2$ decomposition. As can be seen from Fig. 3(b), for the film annealed at 900 °C, two peaks at 1330 and 1620 cm$^{-1}$ were observed, which can be attributed to the D and G bands of carbon materials such as graphite and nano-diamonds. While these modes are barely observable for the sample annealed at 850 °C, they are absent in the films annealed at and below 650 °C. This suggests that carbon contamination has been eliminated by low temperature processing. The presence of carbon also explains the color difference of BaZrS$_3$ processed at different temperatures: while the films processed at 1050 °C show black color (not shown), the films processed at 900 °C and below are dark red.

BaZrS$_3$ thin films annealed at 650 °C and 900 °C were selected for UV-vis absorption measurements to examine their optical properties. The thicknesses of these films are about 280 nm, as measured by AFM (Fig 4(a)). The absorption spectra are shown in Fig 4 (b). It can be seen that their absorption coefficient α above the band gap are ~ 5×10$^4$ and ~ 6×10$^4$ cm$^{-1}$, respectively. These values are somewhat smaller than that of BaZrS$_3$ films processed at higher temperatures [28]. A linear fitting near the band edge of the Tauc plot results in a band gap value of 1.98 and 1.86 eV, respectively, for the films annealed at 650 °C and 900 °C, respectively, as shown in Fig 4(c). These values are slightly higher than the band gap in the films processed at 1050 °C. Several factors could contribute to these differences: 1. The presence of a minute amount of amorphous oxides/oxysulfides, possibly introduced in the target processing, could not be completely eliminated in low temperature processing. The wide gap oxides/oxysulfides may lead to an overestimate of the band gap, as well as lower the absorption coefficient. Careful examination of XRD patterns of samples processed at different temperatures (Fig. 1(a)) reveals no oxide peaks or shift of the peaks compared to the BaZrS$_3$ standard, within the instrument limit. This rules out the presence of measurable amount of crystalline oxides or oxysulfides; 2. The presence of carbon contamination in previous high temperature processed samples may lead to overestimate of the absorption coefficient and underestimate of band gap, as carbon absorbs in a broad range of wavelengths; 3. We observed a small but noticeable shift of the Raman peaks to higher wavenumbers at higher processing temperatures (Fig. S6). While presence of oxysulfides could lead to shift of Raman modes to higher frequencies, this is inconsistent with the fact that this shift is for samples processed at higher temperatures and therefore conversion to the BaZrS$_3$ phase is more complete. We suggest that a slight difference in structural distortion (octahedron rotation) at different processing temperatures is responsible for the observed Raman shift, which could contribute to subtle differences in electronic structures and band gaps [41].

In summary, lower temperature growth of BaZrS$_3$ results in stoichiometric films with minimized concentration of sulfur vacancies, eliminated carbon contamination, smaller crystallite sizes and lower surface roughness. These are expected to strongly affect the carrier transport characteristics and device performance. To investigate the influence of film microstructure on carrier transport, we fabricated and measured photodetector and FET devices using the BaZrS$_3$ films annealed at different temperatures as active semiconductors.

The I-V curves of the photodetector devices were measured in the dark and under illumination, for films annealed at temperatures of 850 °C, 650 °C, and 550 °C, respectively, as shown in Fig 5(a) to 5(c). An optical image of the device is shown in the inset of Fig 5(a). At 850 °C, the I-V curves show little difference under dark and illumination. This suggests that carriers available for the transport are essentially intrinsic to the material, and the contribution from the photo-generated carriers to the transport is negligible. The high dark current can be attributed to the high concentration of sulfur vacancies, as each sulfur vacancy can donate two electrons. Using the sulfur vacancy concentration measured by EDX (Fig S4), we estimate the carrier density to be $1.7\times10^{21}$ cm$^{-3}$. For the film annealed at 650 °C, the measured photocurrent is 80 times greater than the dark current, as can be seen in Fig 5(b). The film annealed at 550 °C shows the highest photocurrent of 17 nA at 10 V. However, the on/off ratio is only 20 times due to the relatively higher dark current. The relatively higher dark current could be caused by the lower crystallinity, where a number of defects may be present and contribute to excess carriers, as shown by our unpublished theoretical study [42]. The photo responsivity is calculated to be 0.08 mA/W at a bias voltage of 5V for the films annealed at 550 °C. The moderate photo response is primarily due to the relatively small grain sizes, which leads to significant carrier scattering and limits the carrier mobility. In Fig 5(d), we plot the photocurrent as a function of time for the films annealed at 650 °C under periodically switched light source at a fixed bias voltage of 8 V. The device shows fast response and low rise/fall time. Due to instrument limit, we estimate the upper limit of the response time to be 0.3 ms. However, the actual response time could be much smaller. The I-V curves measured at different excitation wavelengths for films annealed at 650 °C are plotted in Fig 5(e), and the wavelength dependent photocurrent at bias voltages of 3, 5 and 8 V are shown in Fig 5(f). The device shows the highest photo response at the excitation wavelength around 500 nm. The cutoff wavelength is found to be about 670 nm. This confirms the band gap of the BaZrS$_3$ film annealed at 650 °C to be about 1.85 eV. This suggests that the UV-vis absorption measurements slightly overestimate the band gap, perhaps due to the small amount of oxides.

FET devices were fabricated using the film annealed at 650 °C to further study the carrier transport, using an IL side gate, as shown by the optical image in the inset of Fig 6(a). The drain-source current $I_{DS}$ as a function of bias voltage $V_{DS}$, measured at positive and negative gate voltages $V_G$ are shown in Fig 6(a) and (b), respectively. The total specific capacitance of the IL is determined by the capacitance of the electrical double layer (EDL), and can reach as high as ~20 µF/cm$^2$ [43], which makes it possible to generate high electric field of ~ 10 MV/cm at low gate voltages [44], comparing to commonly used metal oxide dielectric such as SiO$_2$, Al$_2$O$_3$ or HfO$_2$. At a $V_{DS}$ of 1 V, $I_{DS}$ is measured to be $3\times10^{-11}$ A without gating, which suggests very small intrinsic carrier density. The intrinsic carrier type is found to be n-type, which is consistent with our previous work [28]. The intrinsic carrier concentration is found to be $4.4\times10^{10}$ cm$^{-3}$. These results further confirm that carrier concentration is suppressed by reducing the density of sulfur vacancies in BaZrS$_3$ films processed at low temperatures. With increasing $V_G$ to 1V, $I_{DS}$ reaches to $2\times10^{-7}$ A, resulting in an ON/OFF ratio of $6.5\times10^3$. Similar but smaller $I_{DS}$ can be observed for negative $V_G$, as shown in Fig 6(b), suggesting ambipolar behavior in our FETs. We further estimate the electron and hole mobilities to be 16.8 cm$^2$/Vs and 2.6 cm$^2$/Vs, respectively, from the saturated $I_{DS}$ vs. $V_G$ data. The capacitance of the IL layer used in the calculation was obtained from the C-V measurements. The carrier mobility is primarily limited by the small grain sizes which contribute to substantial carrier scattering. Future work should focus on improving the grain size of BaZrS$_3$ films processed at low temperatures to improve the performance of electronic and optoelectronic devices.

**Conclusion**

In conclusion, we have obtained polycrystalline perovskite BaZrS$_3$ thin films, by low temperature crystallization of PLD deposited amorphous BaZrS$_x$ films in CS$_2$. The films with good crystallinity and a surface roughness smaller than 0.6 nm can be obtained at processing temperatures as low as 550 °C. The low processing temperatures result in the elimination of both excessive carriers due to vacancy formation and carbon contamination. Prototype photodetector devices fabricated with such films show good visible light sensitivity, and their wavelength dependent photo response confirms a band gap of ~ 1.85 eV. The FET indicates ambipolarity of the charge carriers, with electron and hole mobilities of 16.8 cm$^2$/Vs and 2.6 cm$^2$/Vs, respectively. Our findings will expedite the progress towards applications of BaZrS$_3$ and other chalcogenide perovskites for electronics and optoelectronics.

**Author contributions**

Z.Y. synthesized the PLD target and thin films, and performed structural and optical characterizations. X.W. fabricated the photodetector and FET devices. X.W. and Y.Z. performed device characterizations and data analysis. H. Z. conceived the idea. S.Y., J.S. and H.Z. supervised the project. X.W., Z.Y. and H.Z. wrote the manuscript. All authors discussed the results and contributed to the final manuscript.

**Acknowledgements**

Work supported by US NSF CBET-1510121, CBET-1510948, MRI- 1229208, and DOE DE-EE0007364. Z. Yu and S. Yang acknowledge NSFC (Grant No. 91963111) and Key Scientific and Technological Innovation Team of Shaanxi Province (2020TD-001). Y. -Y.S. acknowledges support by NSFC under Grant 11774365.

**Conflict of Interest**

The authors declare no conflict of interest.

**Figures captions**

**Figure 1.** (a) XRD patterns of BaZrS$_3$ films annealed at different temperatures. (b) The FWHMs of the (121) peak as a function of annealing temperature. The inset shows the optical images of the BaZrS$_3$ films annealed at the respective temperatures.

**Figure 2**. AFM top-view and 3D topographic images of BaZrS$_3$ thin films annealed at (a) and (e) 900 °C, (b) and (f) 850 °C, (c) and (g) 650 °C, and (d) and (h) 550 °C. (i) The height profile measured along a scan length of 5 µm. (j) The root mean square (RMS) surface roughness R$_q$ as a function of annealing temperature.

**Figure 3**. (a) The room temperature Raman spectra of BaZrS$_3$ films annealed at different temperatures. (b) Raman spectra at higher wavenumbers showing presence/absence of carbon contamination.

**Figure 4.** (a) The AFM height profiles, (b) The absorption coefficient, and (c) The Tauc plot of the films annealed at 650 °C and 900 °C, respectively.

**Figure 5**. The I-V curves in dark and under illumination of the photodetectors fabricated from the films annealed at (a) 850 °C, (b) 650 °C, and (c) 550 °C. The inset in (a) is an optical image of the device. (d) Current vs time for the device using the film annealed at 650 °C, with a periodically switched light source (532 nm) at the fixed bias of 8 V. (e) The I-V curves measured at different excitation wavelengths for the device in (d). (f) The wavelength dependent photocurrent at bias voltages of 3, 5 and 8 V extracted from (e).

**Figure 6**. The I$_{DS}$ vs V$_{DS}$ of the FET fabricated using the film annealed at 650 °C under (a) positive and (b) negative gate voltages V$_G$. The inset in (a) is an optical image of the FET with IL side-gate.

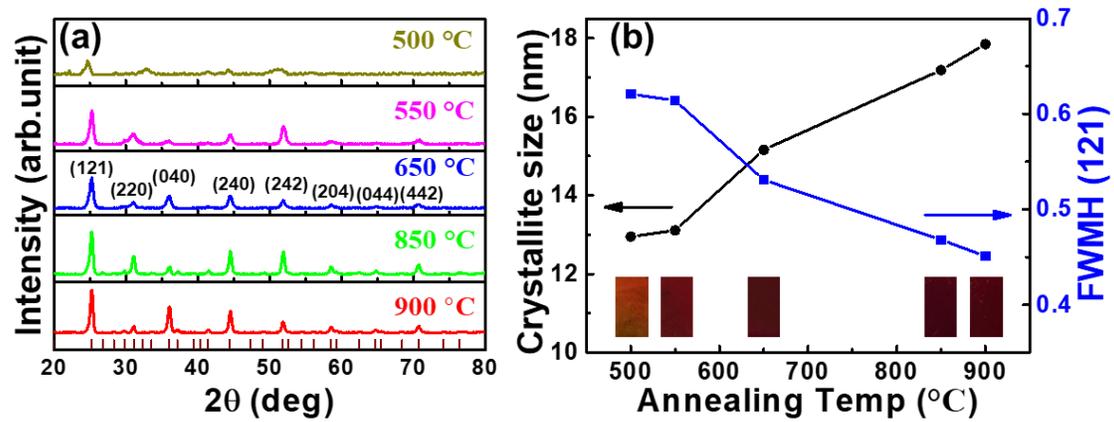

**Figure. 1**

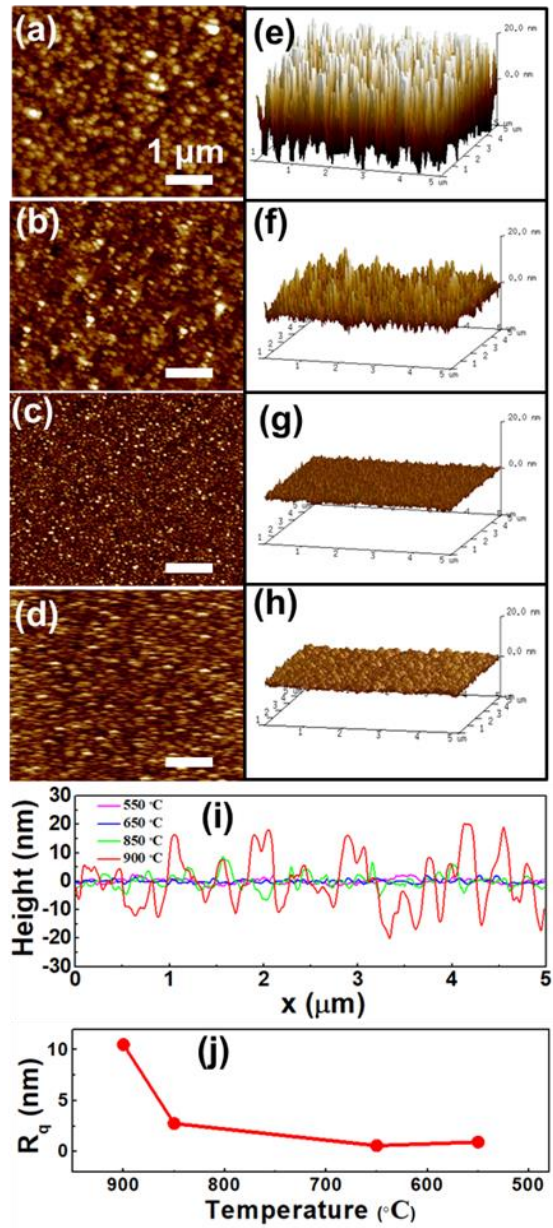

**Figure 2**

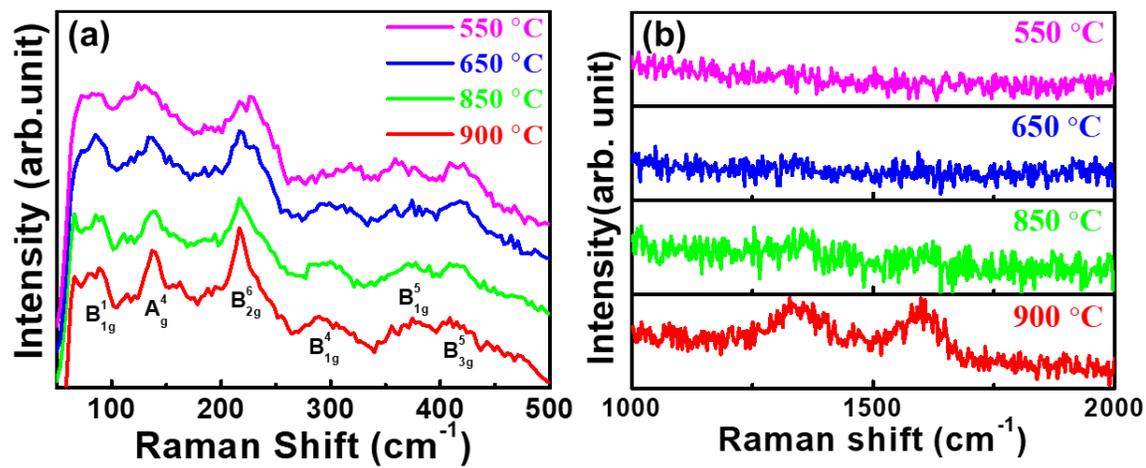

Figure 3

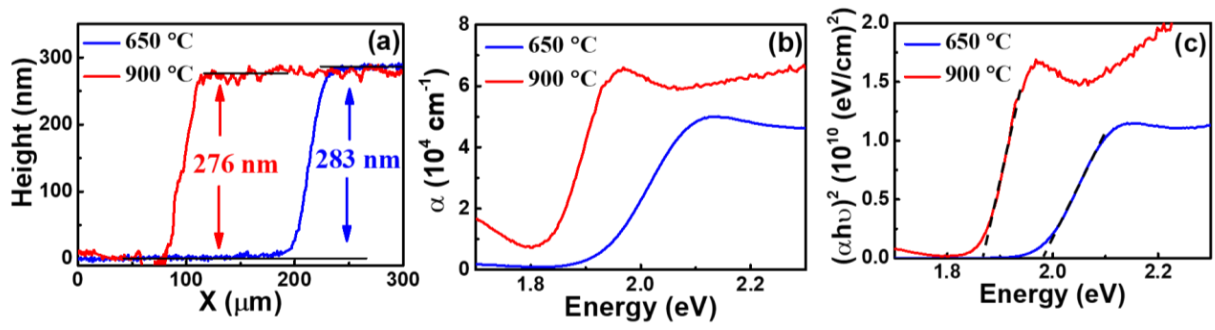

**Figure 4**

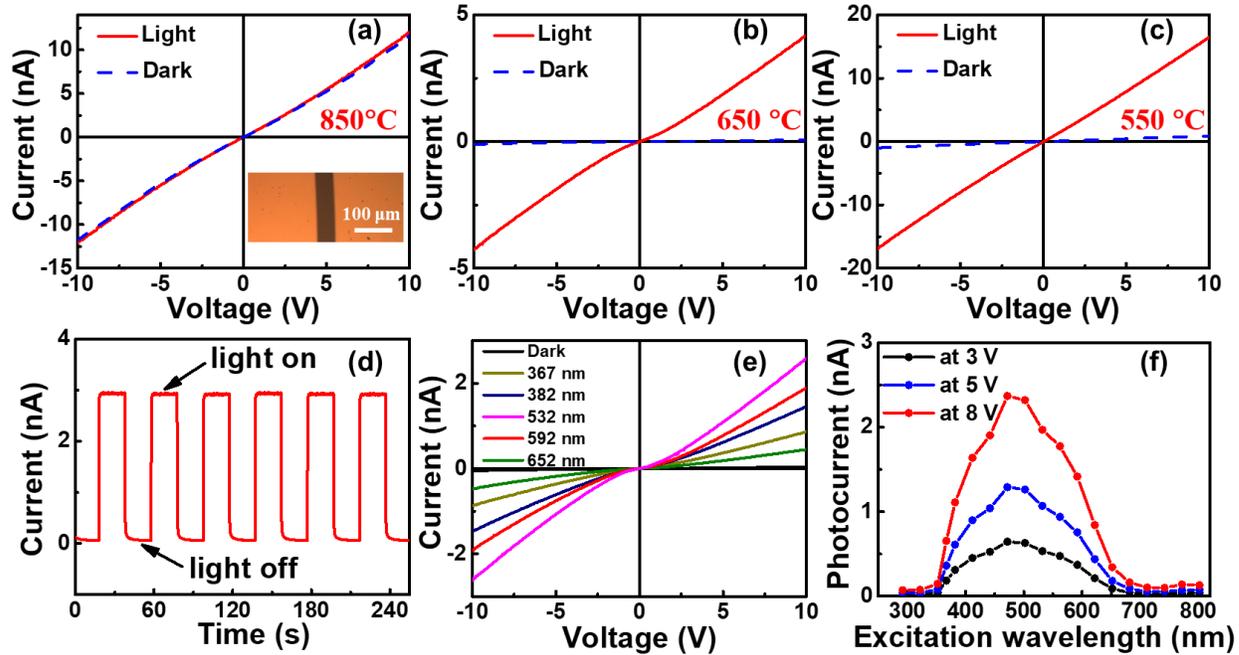

**Figure 5**

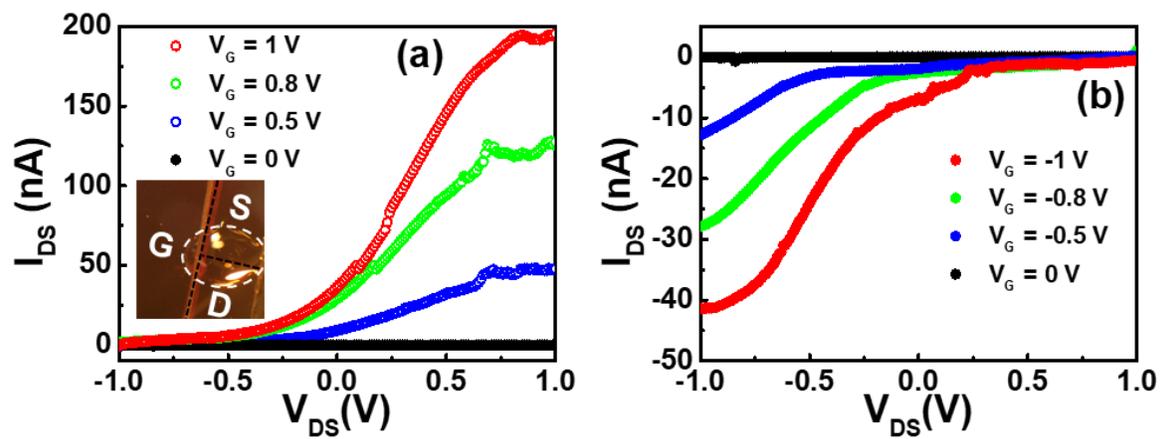

**Figure 6**

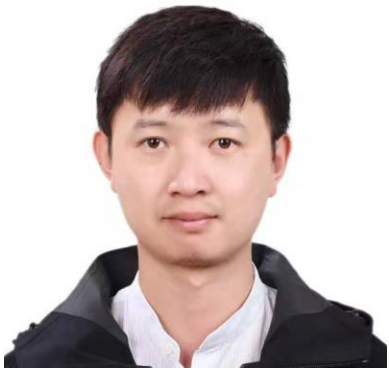

**Zhonghai Yu** received his bachelor of science degree from Taiyuan University of Technology in 2014. He is currently a doctoral student working on his PhD, focusing on Chalcogenide Perovskites solar cell at Xi'an Jiaotong University since 2015. He will be a visiting PhD student at the University at Buffalo.

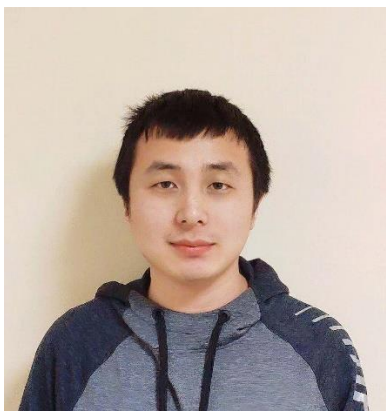

**Xiucheng Wei** received his B.S. degree in physics from Xi'an Jiaotong University, China. He is currently a graduate student in the department of physics at University at Buffalo. His research interests focus mainly on novel photovoltaic semiconducting materials and devices.

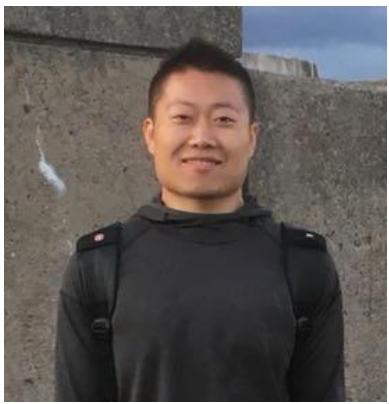

**Yixiong Zheng** is currently pursuing his Ph.D degree in Prof. Jung-Hun Seo's group, at the Department of Materials Design and Innovation, University at Buffalo. He received his B.S. degree in optoelectronics engineering from University of Electronic Science and Technology of China in 2014. His research interest is the design and fabrication of wide band gap semiconductor devices.

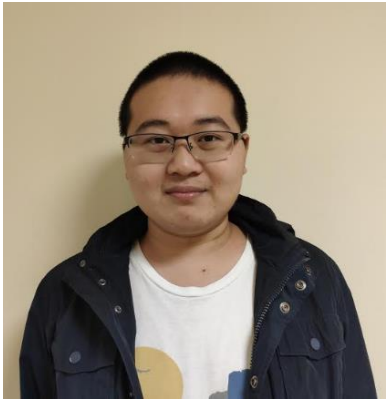

**Haolei Hui** received his bachelor's degree in Xi'an Jiaotong University in 2018. He is currently a PhD student in department of physics at University at Buffalo. He is focusing on novel chalcogenide perovskite as photovoltaic materials.

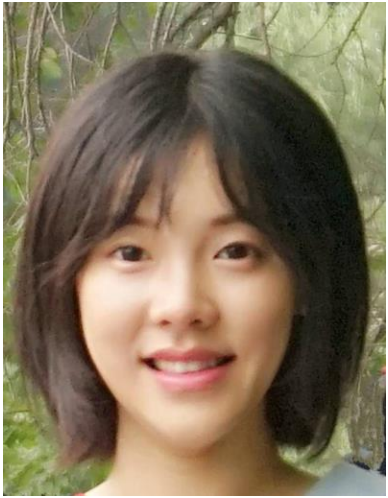

**Mengying Bian** received her Ph.D. in Physics from the Peking University in 2018. Currently, she is working as a postdoctoral researcher in the group of Professor Yanglong Hou at Peking University and a visiting scholar in the group of Professor Hao Zeng at University at Buffalo, where her research is focused on 2D magnetic materials and devices.

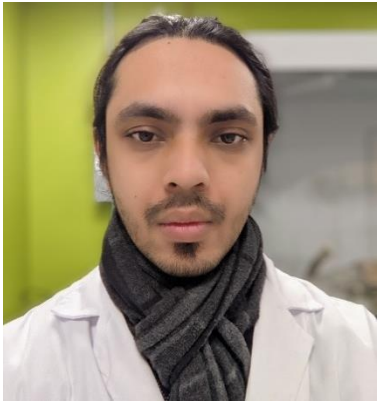

**Samyak Dhole** received his M.Sc. in Chemistry from BITS Pilani, India in 2018. He is currently a graduate student in the Department of Materials Design and Innovation in University at Buffalo. His research interests lie primarily in thin films, synthesis and characterization of perovskite-based materials, resistive switching and polymer assisted deposition.

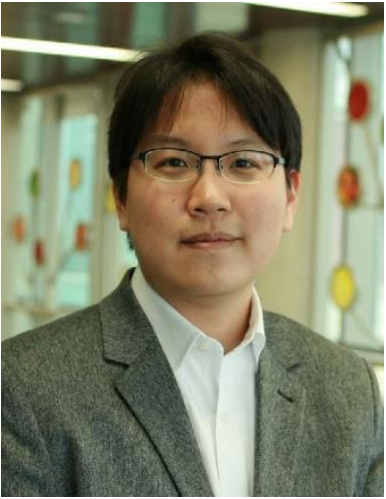

Dr. Jung-Hun Seo received his B.S. degree in electronics and electrical engineering from Korea University, Seoul, Republic of Korea, in 2006. He received his M.S. and Ph.D degrees in Electrical and Computer Engineering from University of Wisconsin-Madison in 2011 and 2014, respectively and worked as an assistant scientist at the University of Wisconsin-Madison from 2014 to 2016. Since 2016, he has been an assistant professor at the Department of Materials Design and Innovation, University at Buffalo. His research interests mainly focus on the synthesis and heterogenous integration of low dimensional wide bandgap semiconductors and their high frequency and high power devices and flexible device applications.

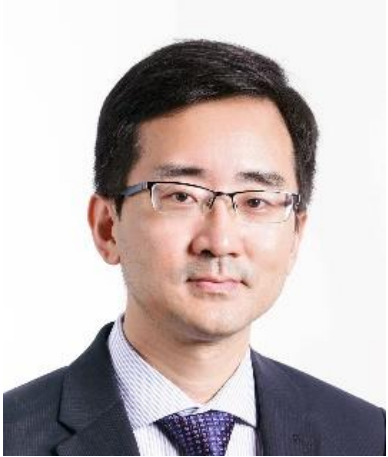

**Yi-Yang Sun** received his Ph.D. in physics from National University of Singapore (NUS) in 2004. Since then, he has worked as a postdoc at NUS, National Renewable Energy Laboratory and Rensselaer Polytechnic Institute (RPI), USA. In 2010, he was appointed Research Assistant Professor and later Research Scientist at RPI. In 2017, he assumed a Professor position at Shanghai Institute of Ceramics, Chinese Academy of Sciences. He has been working on first-principles study of energy-related materials.

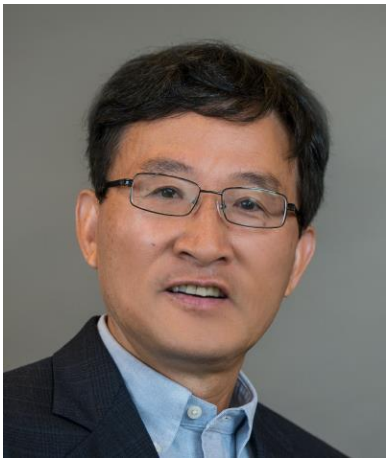

**Quanxi Jia** is an Empire Innovation Professor and National Grid Professor of Materials Research at the University at Buffalo (UB). Prior to joining UB in 2016, he had worked at Los Alamos National Laboratory for 24 years, with the last two years serving as the co-Director and then Director of the Center for Integrated Nanotechnologies, a US Department of Energy Nanoscale Science Research Center operated jointly by Los Alamos and Sandia National Laboratories. His research focuses on nanostructured and multifunctional materials, with a particular effort on the synthesis and study of processing-structure–property relationships of epitaxial films for energy and electronic applications.

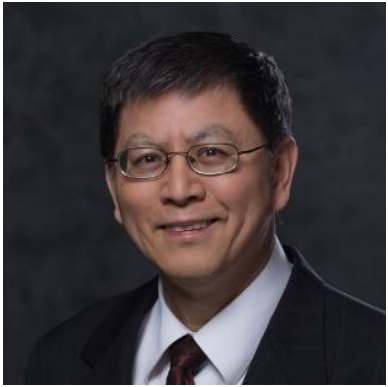

**Shengbai Zhang** received his Ph. D. in physics from University of California at Berkeley in 1989. He moved to Xerox PARC as a postdoc, before joining the National Renewable Energy Laboratory in 1991. In 2008, he became the Senior Kodosky Constellation Chair and Professor in Physics at Rensselaer Polytechnic Institute. His expertise is first-principles theory, modeling, and calculation. Recent work involves chalcogenide perovskites for photovoltaic, ultrafast phase change memory materials, topological carbon networks, non-equilibrium growth, excited state dynamics, and unconventional two-dimensional materials and excitonic insulators. He is a Fellow of the American Physical Society since 2001.

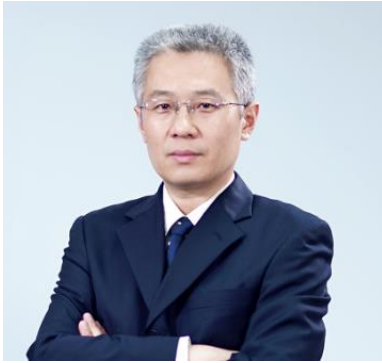

**Sen Yang** received his Ph.D. in materials physics from the Xi'an Jiaotong University (XJTU), China in 2005. He joined the National Institute for Materials Science, Japan in 2005 as a JSPS (Japan Society for the Promotion of Science) post-doctor. In the year of 2010, he came back to XJTU and was promoted to full professor in 2013. His research interests are in magnetism and magnetic materials, smart materials, phase transition and so on.

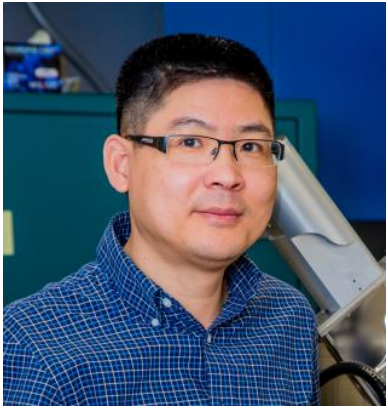

**Hao Zeng** received his Ph.D. in physics from the University of Nebraska-Lincoln in 2001. He joined the Department of Physics at the University at Buffalo in 2004 after a 3-year postdoc at IBM T.J. Watson Research Center, and was promoted to full professor in 2014. His research interests are in nanoscale magnetism and magnetic materials, spintronics, 2D materials and unconventional semiconductor materials and devices.